\documentclass[twoside,leqno,twocolumn]{article}

\usepackage[letterpaper]{geometry}

\usepackage{ltexpprt}
\usepackage{hyperref}

\usepackage{graphicx}
\usepackage{bm}
\usepackage{multirow}
\usepackage{algorithm}
\usepackage{algorithmic}
\usepackage{booktabs}
\usepackage{amsfonts}
\usepackage{amsmath}
\usepackage{subfigure}
\usepackage{adjustbox}
\usepackage{setspace}
\usepackage[font=footnotesize]{caption} 
\begin{document}

\title{\Large Towards More Robust and Accurate Sequential Recommendation with Cascade-guided Adversarial Training}

\renewcommand*{\thefootnote}{\fnsymbol{footnote}}

\author{Juntao Tan\footnotemark[1] \footnotemark[2]\hspace{-0.5cm} \and Shelby Heinecke \footnotemark[3]\hspace{-0.5cm} \and Zhiwei Liu \footnotemark[3]\hspace{-0.5cm} \and Yongjun Chen \footnotemark[3]\hspace{-0.5cm} \and Yongfeng Zhang\footnotemark[1]\hspace{-0.5cm} \and Huan Wang\footnotemark[3]}


\date{}

\maketitle

\footnotetext[1]{Rutgers University. \\ \{juntao.tan, yongfeng.zhang\}@rutgers.edu}
\footnotetext[2]{This project was completed while participating in an internship at Salesforce Research.}
\footnotetext[3]{Salesforce Research. \{shelby.heinecke, zhiweiliu, yongjun.chen, huan.wang\}@salesforce.com}


\fancyfoot[R]{\scriptsize{Copyright \textcopyright\ 2024 by SIAM\\
Unauthorized reproduction of this article is prohibited}}





\begin{abstract} \small\baselineskip=9pt Sequential recommendation models, models that learn from chronological user-item interactions, outperform traditional recommendation models in many settings. Despite the success of sequential recommendation, their robustness has recently come into question. Two properties unique to the nature of sequential recommendation models may impair their robustness - the cascade effects induced during training and the model's tendency to rely too heavily on temporal information. To address these vulnerabilities, we propose Cascade-guided Adversarial training, a new adversarial training procedure that is specifically designed for sequential recommendation models. Our approach harnesses the intrinsic cascade effects present in sequential modeling to produce strategic adversarial perturbations to item embeddings during training. Experiments on training state-of-the-art sequential models on four public datasets from different domains show that our training approach produces superior model ranking accuracy and superior model robustness to real item replacement perturbations when compared to both standard model training and generic adversarial training. 

\end{abstract}

\section{Introduction}
\begin{figure*}[t]
    \centering
    \includegraphics[width=0.7\linewidth]{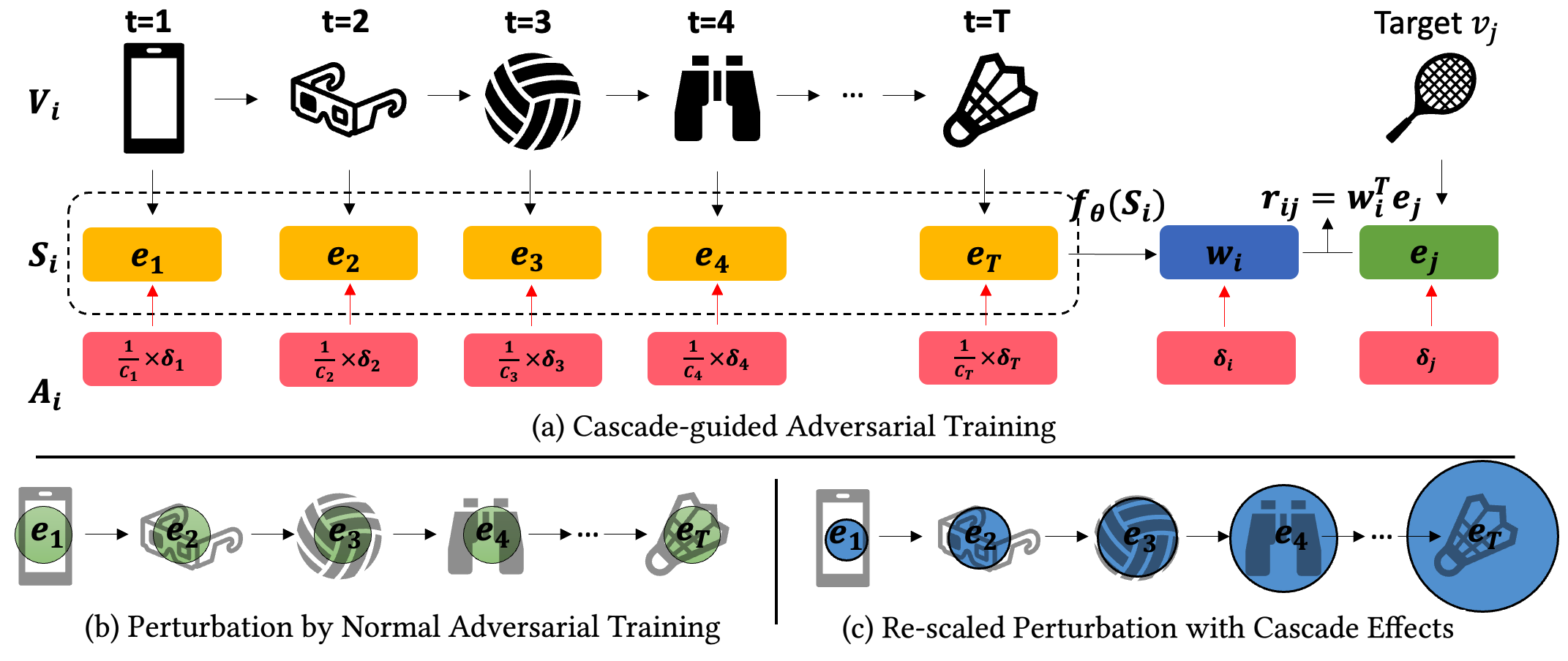}
    \caption{A toy example of applying adversarial training on sequential recommendation. (a) How adversarial perturbations are applied on the learned item and user embeddings. (b) Generic adversarial training applies adversarial perturbations of the same magnitude (green circles) to all item embeddings. (c) The Cascade-guided adversarial training method dynamically choose the magnitude of the perturbations (blue circles) according to the different cascade effects of each interaction in the user history.}
    \label{fig:overview}
\end{figure*}
Sequential recommender models learn dynamic user preferences and recommend next items by modeling past user behaviors in sequential order. Recently, 
deep learning based models have gained attention. These models learn user preferences by feeding the user histories into deep neural networks, such as RNN \cite{hidasi2015session, wu2017recurrent, hidasi2018recurrent}, CNN \cite{tang2018personalized, yan2019cosrec}, or transformer \cite{kang2018self, sun2019bert4rec, fan2021continuous, liu2021augmenting}.

Despite their outstanding performance, recent works reveal their unique robustness issues. A small change at the end of the user sequence, sometimes even on only one item, can dramatically change model behavior \cite{zhang2020practical, yue2021black}. In real-world applications, this kind of instability can lead to serious user dissatisfaction. For example, if a user accidentally clicks on an unwanted video and then returns to the previous page, they might find that all the recommended videos have changed due to that single misclick. In general, robustness is crucial for recommender systems as it affects user trust and acceptance \cite{o2004collaborative, adomavicius2010stability, logesh2020enhancing}. Meanwhile, robustness and accuracy are sometimes correlated \cite{o2004collaborative}, and training a more stable recommender system may also lead to more accurate predictions. Previous work in robustness and stability in recommender systems focus on model families such as matrix factorization \cite{he2018adversarial, park2019adversarial}. Enhancing robustness and accuracy in complex models like sequential recommendation remains relatively untouched.


Sequential recommendation models possess two unique properties that may hinder their robustness: (1) Though they use temporal information to learn user preferences, they tend to \textbf{overemphasize} recent interactions, making the model's recommendations highly sensitive to them \cite{zhang2020practical, yue2021black}. (2) Due to their time-aware training, these models inherently exhibit \emph{cascade effects}, which refers to the fact that changes in a user's behavior can indirectly affect recommendations for others through collaborative filtering \cite{oh2022rank, christakopoulou2019cascade}. In sequential recommender models, item interactions in early timestamps generally induce larger cascade effects, since they affect the predictions on every subsequent interaction in each training epoch \cite{oh2022rank}. During training, their gradients are computed more frequently and are more diverse. As a result, while trained models might be robust against perturbations on the interactions with higher cascade values (i.e., the “early interactions”), they are less robust to perturbations on later ones. Paradoxically, even though the last few user interactions greatly influence predictions, they are underrepresented during training. The combination of these properties exacerbates the vulnerability to perturbations at the end of the user history sequences.


In this study, we enhance robustness by modifying the training process. Noting that interactions with lower cascade effects often play a pivotal role in predictions, we boost stability for these interactions through adding more noise on them during training via adversarial training. Adversarial training has been shown to improve the robustness of a variety of deep learning models by adding adversarial perturbation on the input data \cite{wiyatno2019review}. Originally proposed for image classification, adversarial training produces image classification models that are more robust to adversarial attack on the input pixels \cite{madry2017towards}. 
In the sequential recommendation setting, we expect that by adding greater adversarial perturbations to the interactions with lower cascade effects, the trained model will be more robust to perturbations occurring among the most recent items in the user sequences when making predictions. Based on this, we propose a Cascade-guided Adversarial Training strategy, which is a re-normalized adversarial training method that is specifically designed for deep sequential recommendation by harnessing the intrinsic cascade effects for each interaction. 


Our method distinguishes itself from previous research in two main aspects: (1) While prior studies focus on adversarial training for early recommendation models like matrix factorization and collaborative filtering \cite{deldjoo2021survey}, we explore its application specifically to deep sequential recommendation models, leveraging their unique characteristics; (2) Unless previous works focus on improving generalization or robustness to latent embedding perturbations, instead, we emphasize robustness against adversarial perturbations on real inputs, namely, item interaction sequences, which are more likely perturbations in practical scenarios.

Our work makes the following contributions:
\begin{itemize}
\setlength\itemsep{0pt}
    \item We propose a novel adversarial training algorithm that uses the cascade effects to construct strategic adversarial examples during training.
    \item Through rigorous experiments on four datasets using two widely used sequential recommendation models, we demonstrate our approach outperforms standard and generic adversarial training in terms of accuracy, generalization, and robustness against realistic item replacement perturbations.

\end{itemize}

\section{Related Works}
\subsection{Adversarial Training.}
Adversarial training framework emerged from the seminal paper \cite{szegedy2013intriguing}. 
It demonstrates that images with human-imperceptible perturbations, \textit{i.e.} adversarial examples, could be misclassified by well-trained computer vision classifiers.
Thus, those perturbations also need to be tackled during model training phase, which aims to enhance the robustness of models against such perturbations \cite{wiyatno2019review}. 
Key advancements encompass the Fast Gradient Sign Attack (FGSA) \cite{goodfellow2014explaining} and the min-max optimization formulation \cite{madry2017towards}. 

Adversarial training has recently extended to fields like NLP, using word embedding perturbations \cite{miyato2016adversarial}, and recommendation, with user and item embedding perturbations \cite{he2018adversarial}. However, in recommendation systems, users typically access only item interactions, not model internals, such as the embeddings. We contend that adversarial threats might more authentically come from item misclick-induced interaction perturbations than from direct embedding tweaks. To the best of our knowledge, adversarial training has yet to be explored as a method to enhance robustness against perturbations on model externals.

Existing adversarial training methods for recommendation systems target model types like matrix factorization, collaborative filtering, and others, as reviewed in \cite{deldjoo2021survey}. While one study has attempted adversarial training for sequential recommendation \cite{manotumruksa2020seq}, none have crafted a general approach that capitalizes on the unique characteristics of these models as we do.

\subsection{Robustness of Recommendation}
Robust recommender systems can be categorized into three primary definitions. First, the most prevalent definition centers on the system's accuracy amidst noisy data \cite{o2004collaborative, gao2014robust, shriver2019evaluating}, given that users may inadvertently provide inaccurate ratings. Here, robustness gauges the accuracy shift relative to data noise. Second, robustness hinges on the stability of the recommendations \cite{adomavicius2010stability, adomavicius2011maximizing, adomavicius2016classification, logesh2020enhancing}. Stability in recommendations is evaluated by checking model predictions on unknown items after integrating its own predictions into the training data. A robust model remains largely unchanged under these conditions. Thirdly, robustness is viewed through an attack-defense lens \cite{deldjoo2021survey, yue2021black}. Here, malicious parties might launch attacks on models, for instance, using fake user profiles. Such attacks aim to either push specific items or sabotage the system. Robustness thus signifies the system's ability to fend off these threats. For insights on attack-defense tactics and further robustness aspects in recommendations, refer to \cite{deldjoo2021survey} and \cite{2022ovaisirecsys} respectively.

The key goal of our work is to train robust sequential recommendation models with respect to the first definition of robustness, where our noise model is an adversarial item replacement scheme. We motivate the adversarial item replacement by first showing that replacing later items in interaction histories has a stronger negative impact to ranking accuracy than replacing items earlier in interaction histories. We then develop our novel adversarial training algorithm and show that it enhances resilience to this type of noise and it enhances generalization on unperturbed data.

\section{Preliminary}
In this section, we briefly introduce the basic ideas of adversarial training and sequential recommendation. 
\subsection{Adversarial Training}
Adversarial training is defined as a robust optimization problem with saddle point (min-max) formulation \cite{madry2017towards}. Consider a general classification problem with $d$-dimensional input data $x \in \mathbb{R}^d$ and corresponding label $y \in \mathbb{Z}$ under data distribution $\mathcal{D}$. For a classification model $f_\theta$, the training goal is to minimize the risk $\mathbb{E}_{x,y\sim \mathcal{D}}[L(f_\theta(x), y)]$. 
Adversarial training aims to find a perturbation $\delta$ with bounded norm $||\delta||<\epsilon$ that maximizes the minimum risk with respect to $\theta$.
This can be summarized as the following min-max equation:
\begin{equation}
\small
    \label{eq:pre-adv}
    \min\limits_\theta \left( \max\limits_{\delta, ||\delta||<\epsilon}\mathbb{E}_{x,y\sim \mathcal{D}}[L(f_\theta(x+\delta), y)] \right)
\end{equation}
Fast gradient sign method (FGSM) \cite{goodfellow2014explaining} is the most commonly used method that solves for $\delta$.
FGSM simply generates adversarial perturbations by multiplying the sign of the gradient of the loss function by the maximal perturbation magnitude, $\epsilon$:
\begin{equation}
\small
\label{eq:FGSM}
    \delta = \epsilon \cdot sign \big(\nabla L(f_\theta(x+\delta), y)\big)
\end{equation}



\subsection{Sequential Recommendation}
Our work focuses on adversarial training for sequential recommendation systems. Sequential recommendation systems learn from the ordering of historical user-item interactions to predict the next user-item interaction. We formalize the key components as follows. Let $i \in [1, m]$ denote the index of a user and 
and $\mathcal{V}=\{v_1, v_2, \cdots, v_n\}$ denote the set of all possible items. Suppose user $i$ has a sequentially ordered interaction history of length $T$, $\mathcal{V}_i = \{v^{t}_i \mid t=1, \cdots, T\}$. In practice, each user has different length of history. $T$ is a hyper parameter that decides the maximum length of user history, such that only the last $T$ interactions is considered when making predictions. A sequential recommendation model learns the embedding for each item $v_i$ denoted as $\bm{e_i}$. Together, these embeddings form the item embedding matrix, $\bm{E}\in \mathbb{R}^{n\times d}$. For each user $i$, we concatenate the sequence of embeddings of items in $\mathcal{V}_i$, denoted as $\bm{S_i}=[\bm{e}^{t}_i \mid t=1, \cdots, T]$. The sequential models learn the user embedding, $\bm{w}_i$, as a function of the sequence embeddings,
\begin{equation}
\small
    \bm{w_i} = f(\bm{S_i};\theta),
\end{equation}
where $f$ denotes a sequence embedding model, such as Transformer, and $\theta$ denotes the model parameters. After learning $\bm{w}_i$, for a target item $v_j$, the ranking score $r_{i,j}$ is predicted by
\begin{equation}
\small
    r_{i,j} = \bm{w_i}^T\bm{e_j}.
\end{equation}
During model training, for each user $i$ with target item $v_j$ and a set of negative samples $\mathcal{N}^-\subset \mathcal{V}\setminus v_j$, we minimize the binary cross entropy (BCE) loss defined as:
\begin{equation}
\small
    L_{B}(i,j,\mathcal{N}^-;\theta, \bm{E})=-\big(\log (\sigma(r_{i,j})) + \sum \limits_{v_n \in \mathcal{N}^-}\log (1-\sigma(r_{i,n})\big)
\end{equation}
During training, following the same setting as in \cite{kang2018self}, we truncate the user sequence $\mathcal{V}_i$ according to each timestamp $t$. For each sub-sequence, item $v_i^t$ is treated as the target item and the ranking score is predicted by taking previous items as dynamic user history. Meanwhile, only one negative item $v_n$ is sampled for each $v_i^t$ in each sub-sequence. For simplicity, in the rest of the paper, we denote the ranking loss for item $v_j$ to user $i$ with negative sample $v_n$ as $L_{\text{B}}(i,j,n)$.


\section{Method}
We first introduce the proposed adversarial training algorithm that considers the cascade effects in sequential recommendation. Then, we propose an algorithm to effectively compute the cascade effects for each interaction in the user histories.
\subsection{Cascade-guided Adversarial Training}
Since recommender systems take discrete input data (i.e., user/item IDs), adversarial perturbations are commonly applied on the latent embeddings. As shown in Figure \ref{fig:overview} (a), a deep sequential recommendation model is a hierarchy consisting of two levels: (1) a non-linear deep neural network that takes the sequence embeddings of the user's history $\bm{S_i}$ as input and outputs the user embedding $\bm{w_i}$; (2) a linear model that predicts the final ranking score for user $i$ on target item $v_j$ by linearly multiplying their embeddings as $\bm{w_i^T}\bm{e_j}$. We apply adversarial training on both levels separately. 

For the first level, the adversarial training objective is described as follows. When perturbing the item embeddings in the history sequence within a certain small magnitude, even in the worst case, the learned user embedding shouldn't be too different from the original prediction. Suppose for user $i$ with sequence embeddings $\bm{S_i}$, we denote the adversarial perturbations on the sequence embeddings as $\bm{A_i}=[\bm{\delta_i}^{t} \mid t=1,\cdots,T]$, where $\bm{\delta_i}^{t}$ is the perturbation applied on the corresponding item embedding $\bm{e_i^t}$. We mathematically formulate this objective by an adversarial loss function:

\begin{equation}
\small
    \label{eq:adv1}
    \begin{aligned}
    &L_{\text{adv-1}}(i, \bm{A_i})=||\bm{\hat{w}_i} - \bm{w_i}||_2 \\
    &\text{where} ~~ \bm{w_i}=f(\bm{S_i};\theta), \;\;\;\;\bm{\hat{w}_i}=f\big(\bm{S_i} + \frac{1}{\bm{C_i}}\odot \bm{A}_i;\theta \big)\\
\end{aligned}
\end{equation}
In Eq \eqref{eq:adv1}, the vector $\bm{C_i} \in [1, +\infty)^T$ denotes the cascade effects of each interaction in user $i$'s history. 
Higher values of $\bm{C_i}$ denote higher cascade effects. The way to calculate these cascade effects will be introduced in the next section. The factor $\frac{1}{\bm{C_i}}$ plays a crucial rule in our method: it re-scales the adversarial perturbations so that interactions with smaller cascade effects will receive a larger adversarial perturbation. During sequential recommendation model training, interactions with larger cascade effects are used more often in training than interactions with smaller cascade effects \cite{oh2022rank}, hence, the latter can be more vulnerable and unstable (justified later in Section \ref{sec:exp2}). By applying larger perturbations on the interactions with lower cascade effects, we obtain a model that is more equally robust across all sequence embeddings.

To approximate the worst case adversarial perturbation $\bm{A_i}$, we apply the FGSM attack introduced in Eq.\eqref{eq:FGSM}:
\begin{equation}
\small
\label{eq:FGSM_1}
    \mathcal{A}_i= \epsilon \frac{\bm{g}}{||\bm{g}||_2} \;\;\text{where} ~\bm{g}=\frac{\partial L_{\text{adv-1}}(i, \bm{A_i})}{\partial\bm{S}_i}
\end{equation}
We note that $\frac{\bm{g}}{||\bm{g}||_2}$ is the sign of the direction of the applied perturbation and $\epsilon$ is a human defined parameter to determine the general magnitude of the adversarial attack. The specific magnitude of the perturbation on each interaction will be re-scaled by their cascade effects as one of the key features of our method.

The adversarial training on the first level guarantees the aggregated user embedding $\bm{w_i}$ is robust to small perturbations on the user history. On the second level, we apply adversarial training on the learned user embedding and the target item embeddings. This assures that when small changes are applied to the user and target item embeddings, the model is still able to generate highly accurate recommendation results. We use BCE loss as the second adversarial training loss, which is exactly the same loss function used to train the base recommendation model. Suppose $\bm{\delta_i}$, $\bm{\delta_j}$, $\bm{\delta_n}$ are the adversarial perturbations applied on the embeddings of the user $i$, target item $v_j$, and negative sampled item $v_n$, respectively. The second adversarial training loss is defined as:
\begin{equation}
\small
\label{eq:adv2}
\begin{aligned}
    &L_{\text{adv-2}}(i, j, n, \bm{\delta_i}, \bm{\delta_j}, \bm{\delta_n})=-\big(\log (\sigma(\hat{r}_{i,j})) + \log (1-\sigma(\hat{r}_{i,n})\big)\\
    &\text{where} ~~ \hat{r}_{i,j}=(\bm{w_i}^T+\bm{\delta_i})(\bm{e}_j+\bm{\delta_j}), \\
    &\;\;\;\;\;\;\;\;\;\;\;\; \hat{r}_{i,n}=(\bm{w_i}^T+\bm{\delta_i})(\bm{e}_n+\bm{\delta_n})
\end{aligned}
\end{equation}
Here $\hat{r}_{i, k}$ is the predicted ranking score for any user $i$ and item $k$ after perturbation. Similarly, we approximately generate the worst case $\bm{\delta_i}$, $\bm{\delta_j}$, and $\bm{\delta_n}$ within maximum magnitude $\epsilon$ by:
\begin{equation}
\label{eq:FGSM_2}
\small
\begin{aligned}
    &\bm{\delta_i}=\epsilon\frac{\bm{h_i}}{||\bm{h_i}||_2}\;\;\text{where} ~\bm{h_i}=\frac{\partial L_{\text{adv-2}}(i, j, n, \bm{\delta_i}, \bm{\delta_j}, \bm{\delta_n})}{\partial\bm{e_i}}\\
    &\bm{\delta_j}=\epsilon\frac{\bm{h_j}}{||\bm{h_j}||_2}\;\;\text{where} ~\bm{h_j}=\frac{\partial L_{\text{adv-2}}(i, j, n, \bm{\delta_i}, \bm{\delta_j}, \bm{\delta_n})}{\partial\bm{e_j}}\\
    &\bm{\delta_n}=\epsilon\frac{\bm{h_n}}{||\bm{h_n}||_2}\;\;\text{where} ~\bm{h_n}=\frac{\partial L_{\text{adv-2}}(i, j, n, \bm{\delta_i}, \bm{\delta_j}, \bm{\delta_n})}{\partial\bm{e_n}}
\end{aligned}
\end{equation}

The two adversarial training objectives synergistically improve the robustness of the trained model with respect to all the components. Finally, we optimize the model parameters by minimizing the sum of original BCE loss and the two adversarial training losses:
\begin{equation}
\small
\label{eq:cas_adv}
\begin{split}
    \min\limits_{\theta, \bm{E}} L = & L_{\text{B}}(i, j, n) + \lambda_1 L_{\text{adv-1}}(i, \bm{A_i}) \\
    & + \lambda_2 L_{\text{adv-2}}(i,j,n, \bm{\delta_i}, \bm{\delta_j}, \bm{\delta_n})
\end{split}
\end{equation}
It's worth noting that since the cascade effects only affect the model after the general training process, the adversarial training should be applied after the training of the base model. In the adversarial training phase, by minimizing Eq. \eqref{eq:cas_adv}, we expect to learn better item embeddings $\bm{E}$ and the model parameters $\theta$ such that the model is more accurate and robust.

\subsection{Cascade Effects Calculation}
In this section, we will introduce how we calculate the cascade effect matrix $\bm{C} \in [0, 1]^{m\times T}$ for all the interactions in the user histories, in a time efficient manner.

\begin{figure}[t]
    \centering
    \includegraphics[width=0.85\linewidth]{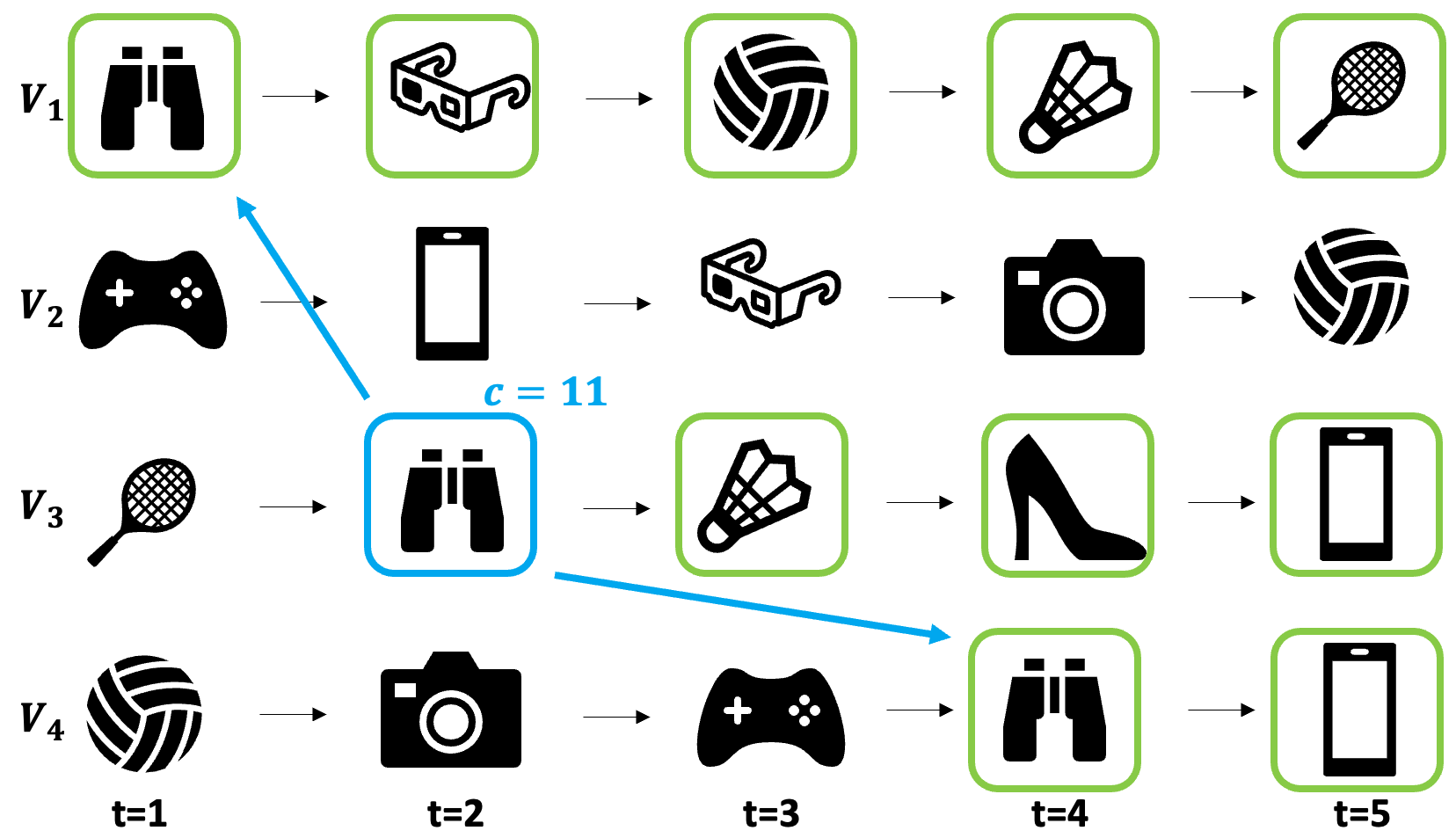}
    \caption{An example of calculating cascade effects. The item with blue bounding box has cascade effects on the $10$ items with green bounding boxes plus itself. Its cascade value is $11$.}
    \label{fig:cascade}
\end{figure}
When training a sequential recommendation model, each user-item interaction produces cascade effects. For a given user-item interaction, two types of interactions receive its cascade effects: (1) all interactions following the given interaction, within the same user history sequence; (2) all interactions with the same item occurring in different user history sequences within the same training batch. This is illustrated in Figure \ref{fig:cascade}.


Based on the above observation, for an item $v_t^i$ for which user $i$ interacted at timestamp $t$, we define the cascade effect $C(i, t)$ as:
\begin{equation}
\small
\label{eq:cascade}
    C(i, t) = 1 + T - t + \frac{b}{m}\sum_{k\in\mathcal{U}, k\neq i}\sum_{l\leq T}(1 + T - l)\mathbb{I}(v_i^{t}, v_k^{l})
\end{equation}
\begin{equation}
\small
    \mathbb{I}(v_{i}^t, v_{k}^l)=    \begin{cases}
      1, & \text{if} ~ v_i^t=v_k^l\\
      0, & \text{otherwise}
    \end{cases}   
\end{equation}
Here, $b$ is the batch size during training and $\frac{b}{m}$ approximates the probability of two user sequences appearing in the same training batch. After calculating $\bm{C}$, its inverse will be a real number in (0, 1], and this will be used to re-normalize the magnitudes of adversarial perturbations.

We note that in Eq. \eqref{eq:cascade}, $1+T-t$ calculates the cascade effects that directly come from the temporal information in the \textbf{same} user history, i.e., the inverse of timestamp. The accumulative term calculates the cascade effects that comes from the same item in \textbf{different} user sequences. In Section \ref{sec:exp1}, we do ablation study to show how much of each type of cascade effects contributes to the final performance.

See supplementary material for a more detailed algorithm for calculating cascade scores.

\section{Experiments}

 We conduct experiments to evaluate the generalization and robustness of deep sequential models trained under our proposed method. Specifically, we consider the following questions:
\begin{itemize}
\setlength\itemsep{0pt}
    \item \textbf{EXP1:} Does the proposed method improve the ranking accuracy of deep sequential recommendation models? How does it compare to normal adversarial training?
    \item \textbf{EXP2:} Does the proposed method improve the robustness of the trained models? If so, which aspect of robustness does it improve, and how much is the improvement?   
\end{itemize}
We first introduce the datasets, base recommendation models, implementation details, and evaluation metrics used in the experiments. Then, in Section \ref{sec:exp1}, we discuss the experiments on ranking accuracy. In Section \ref{sec:exp2}, we address the experiments on robustness. The time efficiency of the proposed adversarial training algorithm is analyzed in the supplementary materials.

\subsection{Datasets}
\label{sec:dataset}
\begin{table}[]
\centering
\caption{Statistics of the datasets.}
\label{tab:datasets}
\begin{adjustbox}{width=0.75\linewidth}
\begin{tabular}{ccccc}
\toprule
Dataset & \#User & \#Item & \#Interaction & Density \\ \hline
ML-1M  &  6,040      &  3,416      & 987,540          &  4.786\%        \\ 
Beauty    & 22,363       & 10,121       & 198,502              & 0.088\%        \\ 
Video  &  24,303      & 10,672       &  231,780             & 0.089\%        \\ 
Clothing  & 39,387   & 23,033 & 278,677 & 0.031\% \\
\bottomrule
\end{tabular}
\end{adjustbox}
\end{table}
\begin{table*}[]
\centering
\caption{Improvement of the accuracy by applying different adversarial training methods on the base models.}
\label{tab:acc}
\begin{adjustbox}{width=0.75\linewidth}
\begin{tabular}{ll|llll|llll}
\toprule
\multicolumn{2}{l|}{\multirow{2}{*}{}}                      & \multicolumn{4}{l|}{MovieLens-1M}        & \multicolumn{4}{l}{Beauty}              \\
\multicolumn{2}{l|}{}                                       & NDCG@10 & vs. Base & HT@10 & vs. Base & NDCG@10 & vs. Base & HT@10 & vs. Base \\ \hline
\multicolumn{1}{l|}{\multirow{5}{*}{SASRec}}  & base        & 0.1359       &  -          & 0.2528      & -           & 0.0264       & -           &    0.0537   &    -        \\ \cline{2-10} 
\multicolumn{1}{l|}{}                         & adv\_linear\cite{he2018adversarial} & 0.1508       & 10.96\%           & 0.2795      & 10.56\%           & \underline{0.0315}       & \underline{19.32\%}           & \underline{0.0626}      & \underline{16.57\%}           \\
\multicolumn{1}{l|}{}                         & adv\_seq\cite{miyato2016adversarial, manotumruksa2020seq}    & 0.1440       & 5.96\%           & 0.2680      & 6.01\%           & 0.0304       & 15.15\%           & 0.0597      & 11.17\%           \\
\multicolumn{1}{l|}{}                         & adv\_global         & \underline{0.1519}       & \underline{11.77\%}           & \underline{0.2808}      & \underline{11.08\%}           & 0.0313       & 18.56\%           & 0.0622      & 15.83\%           \\
\multicolumn{1}{l|}{}                         & \textbf{adv\_cas}    & \textbf{0.1546}       &  \textbf{13.76\%}          & \textbf{0.2831}      & \textbf{11.99\%}           & \textbf{0.0320}       &  \textbf{21.21\%}          & \textbf{0.0630}      & \textbf{17.32\%}           \\ \hline
\multicolumn{1}{l|}{\multirow{5}{*}{GRU4Rec}} & base        & 0.1255       & -           & 0.2419      &  -          & 0.0228       & -           & 0.0460      & -           \\ \cline{2-10} 
\multicolumn{1}{l|}{}                         & adv\_linear\cite{he2018adversarial} & \underline{0.1351}       & \underline{7.65\%}           & \underline{0.2581}      & \underline{6.70\%}           & \underline{0.0288}       & \underline{26.32\%}           & \underline{0.0560}      & \underline{21.74\%}           \\
\multicolumn{1}{l|}{}                         & adv\_seq\cite{miyato2016adversarial, manotumruksa2020seq}    & 0.1308       & 4.22\%           & 0.2520      & 4.18\%           & 0.0252       & 10.53\%           & 0.0493      & 7.17\%           \\
\multicolumn{1}{l|}{}                         & adv\_global         & 0.1345      & 7.17\%           & 0.2543      & 5.13\%        & 0.0285       & 25.00\%           & 0.0550      & 19.57\%           \\
\multicolumn{1}{l|}{}                         & \textbf{adv\_cas}    & \textbf{0.1360}       & \textbf{8.37\%}           & \textbf{0.2588}      & \textbf{6.99\%}           & \textbf{0.0298}       & \textbf{30.70\%}           & \textbf{0.0566}      & \textbf{23.04\%}           \\ \hline
\multicolumn{2}{l|}{\multirow{2}{*}{}}                      & \multicolumn{4}{l|}{Video}               & \multicolumn{4}{l}{Clothing}                \\
\multicolumn{2}{l|}{}                                       & NDCG@10 & vs. Base & HT@10 & vs. Base & NDCG@10 & vs. Base & HT@10 & vs. Base \\ \hline
\multicolumn{1}{l|}{\multirow{5}{*}{SASRec}}  & base        & 0.0441       & -           & 0.0875      & -           & 0.0088       & -           & 0.0184      & -           \\ \cline{2-10} 
\multicolumn{1}{l|}{}                         & adv\_linear\cite{he2018adversarial} & 0.0553       & 25.40\%           & 0.1059      & 21.03\%          & 0.0074       & -15.91\%          & 0.0154      & -16.30\%           \\
\multicolumn{1}{l|}{}                         & adv\_seq\cite{miyato2016adversarial, manotumruksa2020seq}    & 0.0519       & 17.69\%           & 0.1001      & 14.40\%           & \underline{0.0106}       & \underline{20.45\%}           & 0.0213      & 15.76\%           \\
\multicolumn{1}{l|}{}                         & adv\_global        & \underline{0.0557}      & \underline{26.30\%}           & \underline{0.1068}      & \underline{22.06\%}           & 0.0103       & 17.05\%           & \underline{0.0216}      & \underline{17.39\%}          \\
\multicolumn{1}{l|}{}                         & \textbf{adv\_cas}    & \textbf{0.0606}       & \textbf{37.41\%}           & \textbf{0.1162}      & \textbf{32.80\%}          & \textbf{0.0107}       & \textbf{21.59\%}           & \textbf{0.0219}      & \textbf{19.02\%}           \\ \hline
\multicolumn{1}{l|}{\multirow{5}{*}{GRU4Rec}} & base        & 0.0442       & -           & 0.0872      & -           & 0.0072       & -        & 0.0150      & -           \\ \cline{2-10} 
\multicolumn{1}{l|}{}                         & adv\_linear\cite{he2018adversarial} & \underline{0.0500}       & \underline{13.12\%}           & 0.0962      &  10.32\%          & 0.0070       & -2.78\%           & \underline{0.0154}      & \underline{2.67\%}           \\
\multicolumn{1}{l|}{}                         & adv\_seq\cite{miyato2016adversarial, manotumruksa2020seq}    & 0.0456       & 3.17\%           & 0.0888      & 1.49\%           & \underline{0.0074}       & \underline{2.78\%}           & 0.0152      & 1.33\%           \\
\multicolumn{1}{l|}{}                         & adv\_global         & 0.0496       & 12.22\%           & \underline{0.0966}      & \underline{10.78\%}          & 0.0071    &  -1.39\%           & 0.0146      &  -2.67\%          \\
\multicolumn{1}{l|}{}                         & \textbf{adv\_cas}    & \textbf{0.0520}       & \textbf{17.65\%}           & \textbf{0.0993}      &  \textbf{13.88\%}          & \textbf{0.0083}       &  \textbf{15.28\%}          &    \textbf{0.0168}   &  \textbf{12.00\%}           \\
\bottomrule
\end{tabular}
\end{adjustbox}
\end{table*}

We conduct experiments on four diverse public datasets spanning various domains and densities, which are MovieLens-1M \footnote{\url{https://files.grouplens.org/datasets/movielens}} \cite{harper2015movielens} and Video, Beauty, Clothing from Amazon review datasets. They are widely used datasets in recommendation research. Dataset statistics are shown in Table \ref{tab:datasets}.


\begin{figure}[t]
\centering
\begin{adjustbox}{width=0.8\linewidth}
\mbox{

    \subfigure[Train SASRec on Video.]{
        \includegraphics[width=0.25\textwidth]{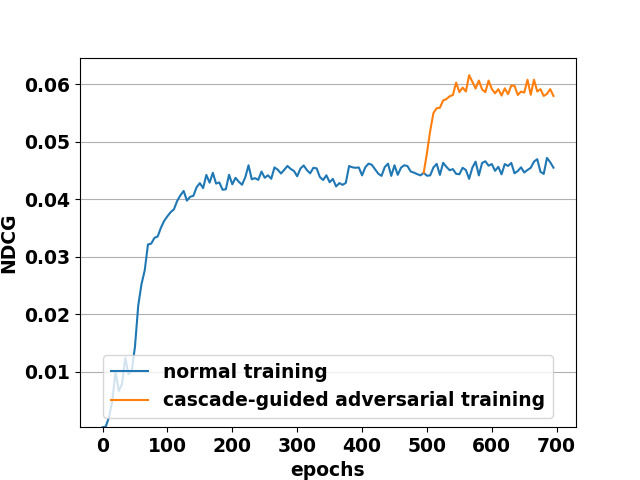}}
    \subfigure[Train GRU4Rec on Video.]{
         \includegraphics[width=0.24\textwidth]{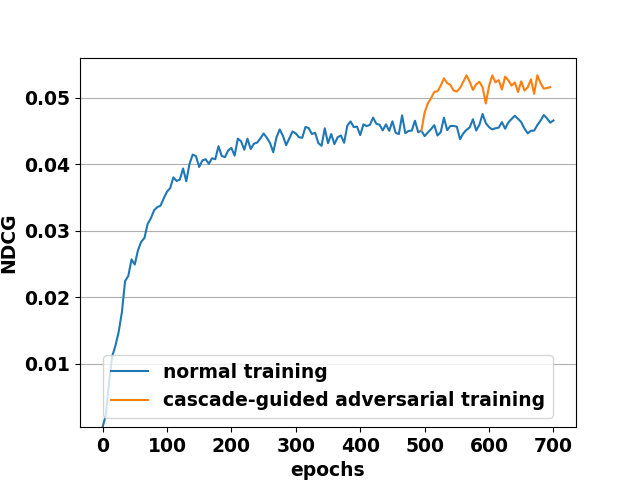}}
}
\end{adjustbox}
\caption{NDCG vs. the number of training epochs. }

\label{fig:learning_curve}
\end{figure}
\subsection{Base Models}
\label{sec:base_models}
Since RNN and transformer are the most common structures been used in deep sequential models \cite{karitarnntransformer}, we choose two of the most representative recommendation models from each of the above categories as the base models: 
\begin{itemize}
    \item GRU4Rec\cite{hidasi2015session}: GRU4Rec utilizes RNNs to learn user preferences based on their history sequences.
    \item SASRec\cite{kang2018self}: SASRec relies on attention mechanisms that can dynamically learn the attention weights on each interaction in the user sequences.
\end{itemize}

\subsection{Implementation Details}
\label{sec:inplementation_details}
First, we implement exactly the same architectures for the two base models as described in their original papers. We use a single layer of GRU units in the GRU4Rec model and $2$ self-attention blocks in the SASRec model. The hidden size is set to $100$ for both models. When computing user embeddings, the maximum sequence length $T$ is set to $200$ for MovieLens-1M and $50$ for Video, Beauty and Clothing datasets according to their different densities. 

We use same training strategy for training all the models on all the datasets: We first train the base models for $500$ epochs to ensure their convergence, then we apply Adversarial Training (generic or our method) on the trained models for further $100$ epochs. For both of the training phases, we use Adam optimizer\cite{kingma2014adam} with $0.001$ learning rate. The batch size is $128$. We use $0.2$ dropout rate and $1 \times 10^{-5}$ $L_2$ norm to prevent overfitting. We follow a leave-one-out strategy to split training and test data, which is commonly used in sequential recommendation.

For the hyper-parameters, the magnitude $\epsilon$ is always set to $10$ for all the datasets. The ablation study on different $\epsilon$ values can be found in Section \ref{sec:exp1}. For parameters $\lambda_1$ and $\lambda_2$ in Eq. \eqref{eq:cas_adv} are always $1$ such that the three loss functions equally contribute to the training. Since no hyper-parameter used to compute the cascade effects, our proposed adversarial training method can be easily applied on new datasets and models without additional tuning effort.

\subsection{Evaluating Metrics}
\label{sec:metric}
We use standard evaluation metrics, Normalized Discounted Cumulative Gain (NDCG) and Hit Ratio (HT), to evaluate the ranking performances of the recommendation models in this paper. Noted that negative sampling is not used in the evaluation. 
Instead, we rank over all items except the ones already in the user histories and retrieve the top-10 ranking list. This evaluation method is used in more and more state-of-the-arts \cite{tang2018personalized, xie2022contrastive}.

\begin{figure}[t]
\centering
\begin{adjustbox}{width=0.85\linewidth}
\mbox{
    \subfigure[SASRec on ML-1M]{
        \includegraphics[width=0.24\textwidth]{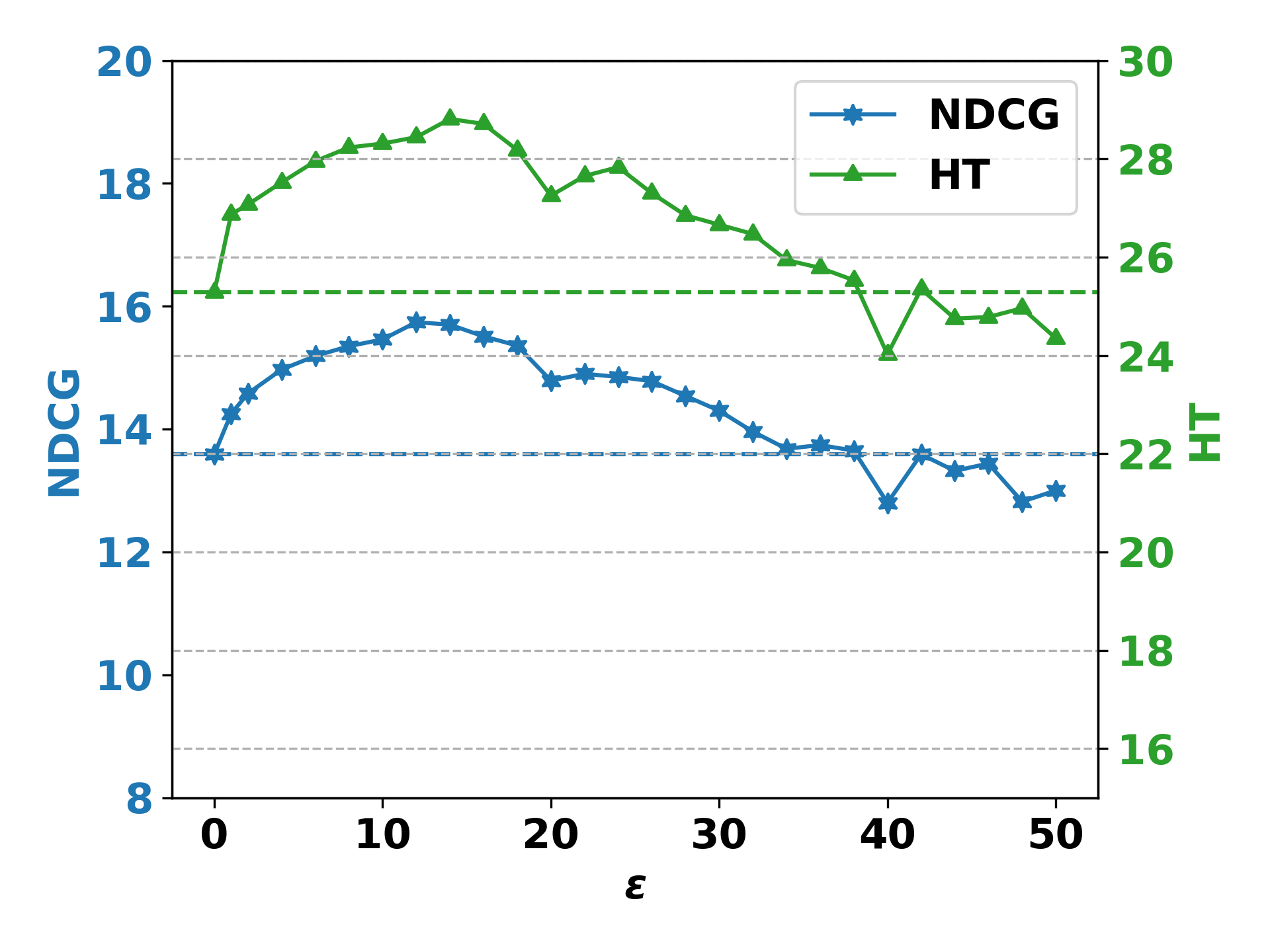}}
    \hspace{-15pt}
    \subfigure[GRU4Rec on ML-1M]{
        \includegraphics[width=0.24\textwidth]{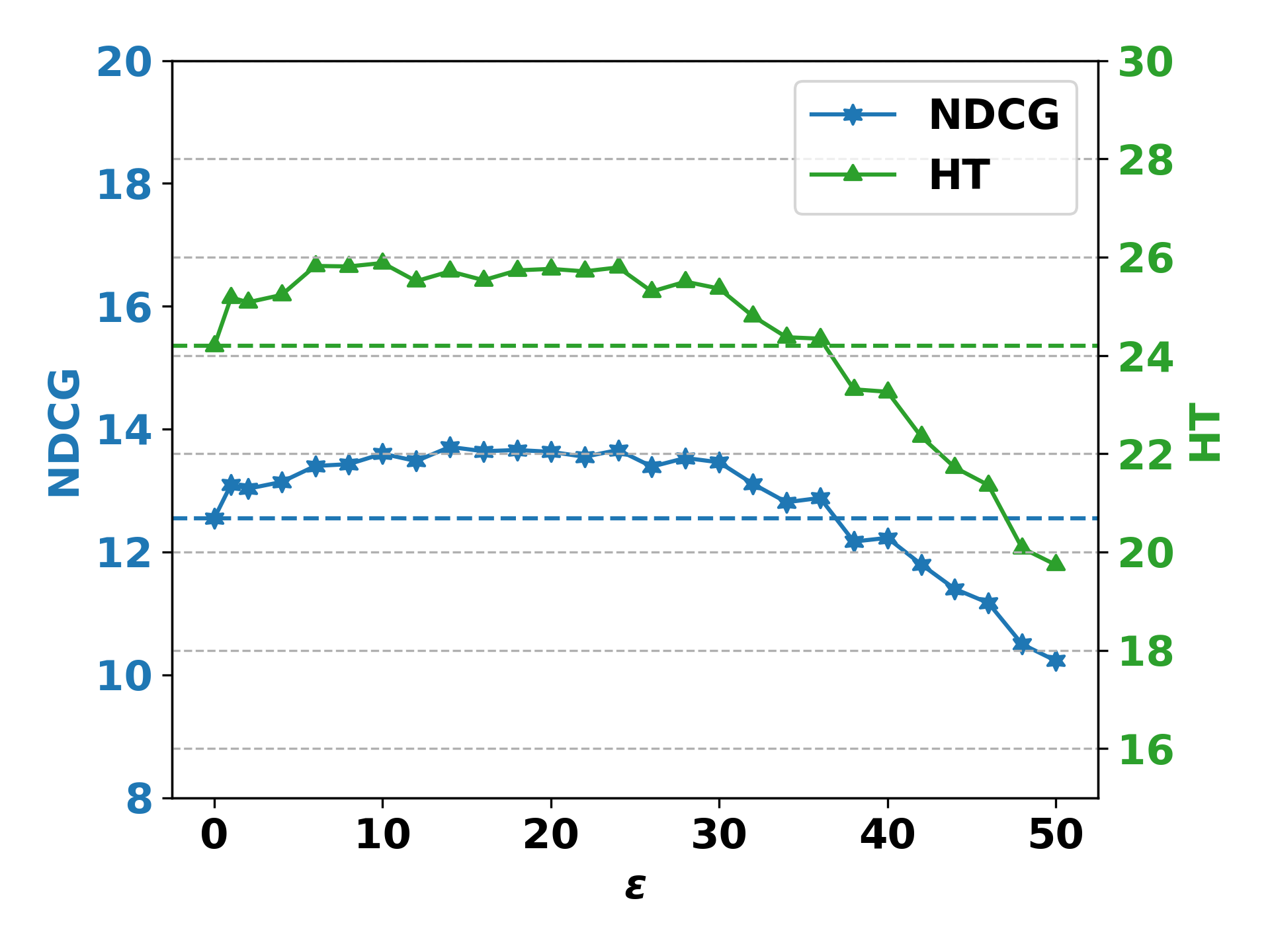}}
}
\end{adjustbox}
\caption{Influence of $\epsilon$}
\label{fig:epsilon}
\end{figure}

\begin{figure}[t]
\centering
\mbox{
\begin{adjustbox}{width=0.85\linewidth}
\centering
    \subfigure[HT@10: SASRec on Video]{
        \includegraphics[width=0.25\textwidth]{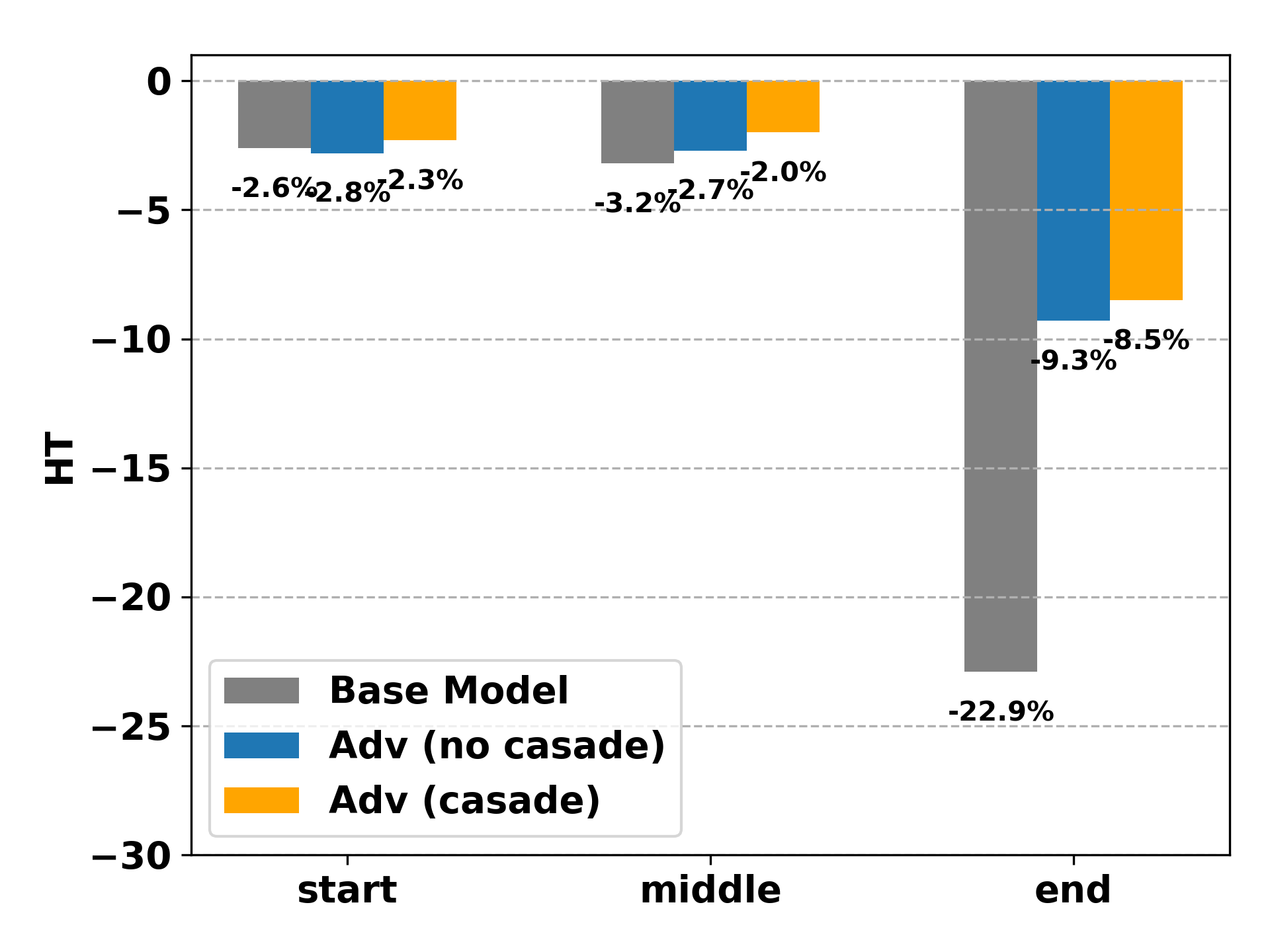}}
    \subfigure[NDCG@10: SASRec on Video]{
        \includegraphics[width=0.25\textwidth]{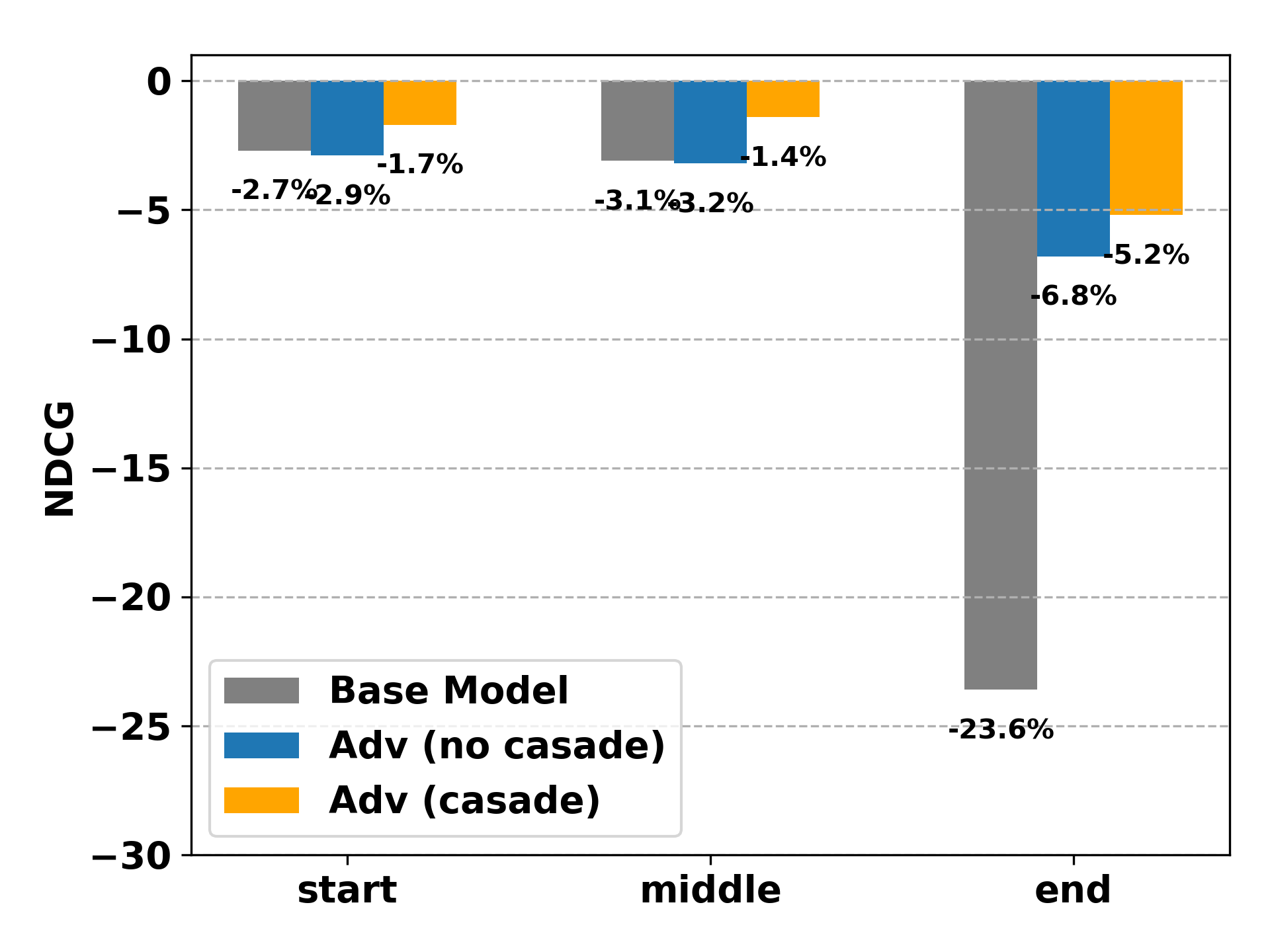}}
\end{adjustbox}
}
\caption{Drop of accuracy by attacking the the first, middle,
and the last items in the user sequences respectively. The y-axis depicts negative values, with larger bars indicating larger decreases in accuracy of the recommendation models.}
\label{fig:robust_drop}
\end{figure}

\begin{figure*}[t]
\centering
\mbox{
\begin{adjustbox}{width=0.85\linewidth}
    \subfigure[SASRec on MovieLen-1M.]{
        \includegraphics[width=0.24\textwidth]{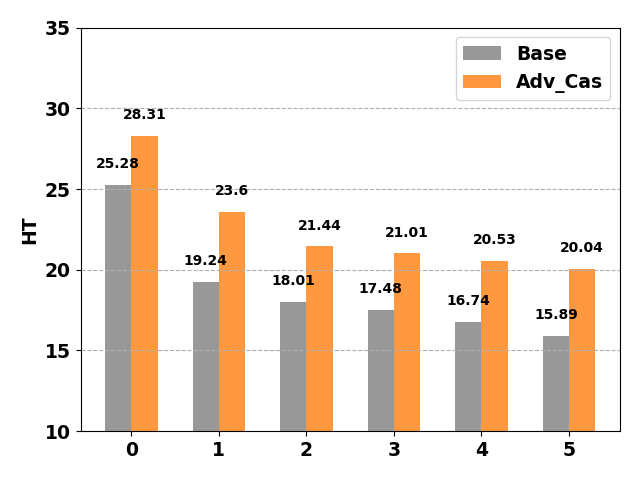}}
    \hspace{1pt}
    \subfigure[SASRec on Video.]{
        \includegraphics[width=0.24\textwidth]{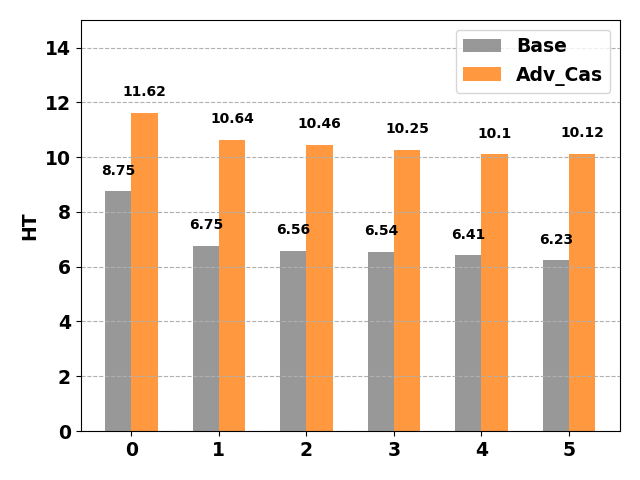}}
    \hspace{1pt}
    \subfigure[SASRec on Beauty.]{
        \includegraphics[width=0.24\textwidth]{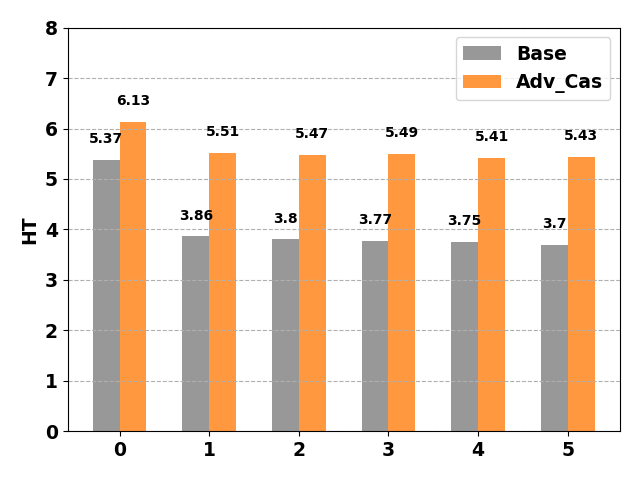}}
    \hspace{1pt}

    \subfigure[SASRec on Clothing.]{
        \includegraphics[width=0.24\textwidth]{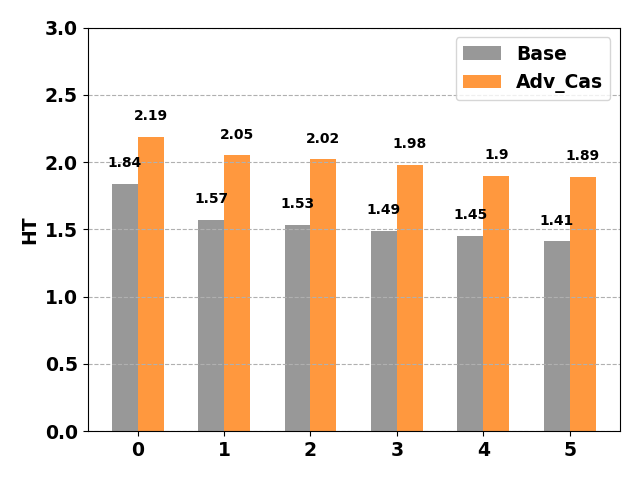}}
\end{adjustbox}
}


\caption{Decreases of model accuracy by replacing K items at the end of user sequences according to HT@10.}
\label{fig:replace}
\end{figure*}

\subsection{EXP1: Improvement on Ranking Accuracy}
\label{sec:exp1}
In the first set of experiments, we evaluate the extent to which our proposed adversarial training method can improve the ranking accuracy in comparison to baseline adversarial training methods. We consider three baseline adversarial training approaches:


\begin{itemize}
    \item \textbf{adv\_linear\cite{he2018adversarial}}: Adversarial training for MF-based recommenders, perturbing only user and item embeddings, i.e., $\bm{w_i}$ and $\bm{e_j}$.
    \item \textbf{adv\_sequence\cite{miyato2016adversarial, manotumruksa2020seq}}: Originally for LSTM text classification\cite{miyato2016adversarial}, later adapted for sequential recommendation\cite{manotumruksa2020seq}, perturbing only sequence embeddings.
    \item \textbf{adv\_global}: Our self-defined baseline, similar to our proposed method but without cascade-guided re-normalization. Serves as an ablation study on cascade value contributions.
\end{itemize}

For fair comparison, when implementing our method, we always re-normalize the average cascade values to $1$ so that the total magnitudes of added perturbations are the same for all the baselines. The experiments results are shown in Table \ref{tab:acc}. First, we observe that compared to the pre-trained base models (no adversarial training), applying any of the adversarial training methods can generally further improve the ranking accuracy. This means that while adversarial training improves the models robustness on the adversarial examples, it also improve the models generalization on the clean data \cite{miyato2016adversarial}. Second, our proposed method consistently outperforms all the other baselines on all the datasets. On average, it improves the ranking accuracy of the base model by $15.32\%$ for SASRec and $15.99\%$ for GRU4Rec. Compared to the second best baselines, it shows $18.16\%$ \textbf{more improvement} of the base model for SASRec and $168.94\%$ for GRU4Rec. Third, the improvements on the sparser datasets (i.e., Video, Beauty, and Clothing) are more significant than the other dataset. This is expected since when training complicated deep neural networks on sparse datasets, the trained models are more easily to overfit on the data and less stable. It's worth noting that all the other adversarial training methods fail to improve GRU4Rec on Clothing dataset possibly due to it's extreme sparsity, but the cascade-guided adversarial training method still performs well.

We illustrate the learning curves of the two training phases in Figure \ref{fig:learning_curve}. As shown in the figures, the base models converge in the first $500$ epochs of pre-training. Training for more epochs can hardly benefits the trained models. However, when applying the Cascade-guided Adversarial Training, the ranking accuracy can be quickly improved further, especially in the next $100$ steps.

We also consider a related question: what should be the magnitude of the adversarial perturbations? The magnitude of adversarial perturbations, controlled by hyper-parameter $\epsilon$ in Eq. \eqref{eq:FGSM_1} and \eqref{eq:FGSM_2}, impacts the model's robustness. Minor perturbations might not enhance robustness, while excessive perturbations can hinder learning. We perform ablation studies on $\epsilon$, by setting its value from $0.1$ to $50$ as shown in Figure \ref{fig:epsilon}. We find our algorithm effective across a spectrum of $\epsilon$. Even with small values like $1$, can significantly improve the models' performance. The method shows the best performance when $\epsilon \in [10, 20]$ for the both base recommendation models. When $\epsilon$ is larger than $30$, the models' accuracy start to drop, suggesting that the adversarial perturbations are too large, and useful information is lost during training.

As in Eq.\eqref{eq:cascade}, each interaction has two types of cascade effects on other interactions (i.e., cascade effects on direct later interactions and cascade effects on the interactions from other sequences).  Please refer to supplementary materials for ablation study about how each component of cascade effects contribute to the final performance.

\subsection{EXP2: Improvement on Robustness}
\label{sec:exp2}
We evaluate the performance of the trained recommendation models in the presence of noise data.  Notably, our evaluations perturb by substituting real items, not merely tweaking item embeddings, therefore, simply increasing the norm of the learned item embeddings can not lead to trivial solutions. 

We ensure replacements are near the original in the embedding space for minimal perturbation. Since major changes in user history should eventually lead to different recommendations, hence, substituting a similar item better mimics ``natural'' data noise, like misclicks. Experimentally, we derive the most adverse minor replacements by first using a gradient-based attack on target item embeddings and then pairing with the nearest items in the gradient direction.

Since the proposed method is motivated based on the idea that recommendation models have varying robustness across a user's history, being more susceptible towards the end. We test this by replacing the first, middle, and last items in user sequences, then measuring accuracy drops. In Figure \ref{fig:robust_drop} using the Video dataset, replacing initial and middle items results in minimal accuracy reduction (under $3\%$). However, changing the last item sees a notable drop ($23.6\%$ in NDCG and $22.9\%$ in Hit ratio). This underscores the models' vulnerability towards sequence ends. Our method significantly improves this end-sequence robustness, outperforming base models and standard adversarial training. 

Next, We evaluate model robustness by replacing the last $K$ items in user sequences. We choose $K$ from $1$ to $5$ to show the trend of decreasing model accuracy. These tests are conducted on both base models across all four datasets. Figure \ref{fig:replace} presents the results for the SASRec model while additional results are in the supplementary materials. Our algorithm-trained models consistently surpass the original models in overall performance, with a slower decline in accuracy at the same time. Notably, for the Video, Beauty, and Clothing datasets, modifying the last $5$ items still results in a ranking accuracy higher than traditionally trained models on clean data, underscoring the efficacy of our method, especially on sparse datasets.


\section{Conclusion}
In this work, we introduce Cascade-guided Adversarial Training, a novel adversarial training algorithm designed for sequential recommender systems. This is the first adversarial training algorithm that considers the intrinsic properties of sequential recommendation models. Using our approach, we show that the trained sequential recommendaion models are more robust and accurate. Considering the small number of additional training epochs and the simplicity of the additional parameters involved in the proposed training process, the proposed method is very practical. 
\bibliographystyle{plain}
\bibliography{main}
\end{document}